# A Local Concentration-based Descriptor Predicting the Stacking Fault Energy of Refractory High Entropy Alloys

Cong Ma[1], Wang Gao[1,*], and Qing Jiang[1,*]

[1]Key Laboratory of Automobile Materials, Ministry of Education, Department of Materials Science and Engineering, Jilin University, 130022, Changchun, China.

*Emails: wgao@jlu.edu.cn; jiangq@jlu.edu.cn.

**Abstract**

Stacking fault energy (SFE) is an essential parameter for characterizing mechanical properties. However, in high entropy alloys (HEAs), the local chemical environment varies significantly across different stacking fault planes, resulting in a substantial fluctuation of SFE values rather than a unique value, which prohibits the prediction of the local SFE. Herein, we proposed an effective descriptor based on the local concentration ratio near stacking fault to quantitatively predict the local SFE of refractory HEAs. We find that the role of a given element in determining SFE strongly depends on its valence-electron number relative to other components and the contribution of its *s*- and *d*-electrons to its cohesive properties, which can be understood in the framework of the tight-binding model. Notably, the descriptor not only unifies the local nature of SFE from simple alloys to HEAs but also helps to quickly design HEAs as the involved parameters are easily accessible.

**Teaser**

A descriptor based on the local concentration ratio near the stacking fault is proposed to predict the stacking fault energies of refractory high entropy alloys.

**Introduction**

Stacking fault (SF), an important surface defect, forms as a result of the dislocation dissociations(*1*, *2*). Therefore, stacking fault energy (SFE), defined as the energy cost per area when forming the stacking fault from perfect crystals, serves as a key parameter for characterizing mechanical properties, such as the dislocation motion, twinning formation, and even phase transition(*3*, *4*). Simply put, a low SFE indicates the twinning formation, while a high SFE favors the dislocation slip(*5–7*). In simple binary alloys, the trend of SFE can be rationalized by the Suzuki mechanism(*8–10*). However, when extended to complex high entropy alloys (HEAs), such as refractory HEAs, the conventional Suzuki mechanism is generally considered invalid(*11*). The complexity of HEAs originating from the random distribution of multi-elements prohibits the establishment of the structure-property relationship of SFE.

Uncovering the intrinsic determinants is the prerequisite to building a predictive descriptor. The attempts can be traced back to the conventional Suzuki mechanism—the aggregation of solute atoms in the SF lowers the SFE of simple alloys(*8–10*), that is the concept of local concentration. The local concentration refers to the compositional fluctuation near the SF planes (that may extend several planes outward)(*5*, *12*). Besides, the SFE of binary alloys also depends on the compositional proportion (i.e., global concentration), such as NiCo(*12*) and FeMn(*13*). Instead of the view of concentration, the electronic descriptors—such as the valence electron concentration (VEC) and charge density redistribution ($\Delta\rho$)—correlate well with SFE but fails quickly in the medium entropy alloys (MEAs) and HEAs due to the average effect of multiple principal elements(*5*, *14*, *15*). For the complex MEAs and HEAs, the models specific to them focus on describing the inhomogeneity of the chemical environment, such as the short-range order (SRO)(*16–18*) and concentration wave (CW)(*19–21*): SRO exhibits infinitesimal concentration fluctuation, correlated with the unlike pair preference, whereas CW is featured with concentration fluctuation deviating from the randomness. For the CrCoNi-based HEAs, both the two contradicting models seem to be able to account for high

SFEs(*16*, *17*, *21*), making the physical picture elusive. This is because both SRO and CW are hard to quantitatively measure in experiments, and thus their existence and effect on the mechanical properties remain debated. Besides, these two models describing the global properties of HEAs are correlated with the statistically-averaged SFE deriving from strikingly different SF planes, instead of relating the local chemical environment of SF with its local individual SFE (i.e., point-to-point prediction). In MEAs and HEAs, the global concentration still serves as a way to tune SFE(*20*, *22–26*). Just like the global SRO and CW, however, the global concentration fails to capture the local nature of SFE and thus fails to establish the point-to-point prediction. The local concentration originating from the conventional Suzuki mechanism has a great advantage in that but disappointedly without quantitative correlation, even in binary alloys (e.g., NiFe ((*5*)) not to mention MEAs (e.g., CrCoNi(*15*) and MoNbTi(*27*)), and HEAs (e.g., CrMnFeCoNi(*15*)).

Herein, we proposed a local concentration-based descriptor to quantitatively predict the SFE of refractory HEAs. A given element behaves substantially different in different HEAs, depending on its valence-electron number relative to other components and its *s*- and *d*-components of cohesive energies. The underlying mechanism could be understood within the tight-binding (TB) framework. Besides, the model with obtainable parameters provides a fast way to screen the HEAs with desirable mechanical properties.

**Results**

The bcc refractory HEAs, an important but less-studied group of HEAs, contain four to nine elements in the compositional space of IV B (i.e., Ti, Zr, and Hf), V B (i.e., V, Nb, and Ta), VI B (i.e., Cr, Mo, and W) and Al elements(*23*, *28–33*). Herein, we adopted 5 quinary HEAs— TiVNbTaMo, TiVNbTaW, ZrVNbTaMo, HfVNbTaMo, and TiNbTaMoW, 6 quaternary HEAs— TiNbTaMo, VNbTaMo, VNbTaW, NbTaMoW, ZrVNbTa, and HfVNbTa, and 2 non-equimolar HEAs—TiNbTaMo2 and TiNbTa2Mo. For the most possible <111>{110} slip system in the bcc crystal, the stacking fault energy refers to the unstable SFE ($\gamma_{usf}$)(*34*). Further details are given in *Methods*.

**The descriptor based on the local concentration** The idea of the local concentration originates from the conventional Suzuki mechanism for simple alloys(*8–10*). After that, the correlation of local concentration to SFE has been extended to MEAs (e.g., CrCoNi(*15*) and MoNbTi(*27*)) and HEAs (e.g., CrMnFeCoNi(*15*)). Different from the general result of the original Suzuki effect in simple alloys, in MEAs and HEAs, the aggregation of a certain component does not always lead to the reduction of SFE but conversely leads to the increase of SFE. Notably, no quantitative relationship between SFE and local concentration has been established, no matter in the simple binary alloys (i.e., original Suzuki effect) and in the complex MEAs and HEAs. For example, the increase of SFE with local Ni content (i.e., Ni aggregation) in the CrCoNi and CrMnFeCoNi shows a highly non-linear behavior(*15*). For the MoNbTi, the correlation refers to the monotonic increase of SFE with Mo aggregation in a certain range(*27*). Instead of correlating with the statistically-averaged SFEs as SRO, CW, and global concentration do, this study aims to relate the local concentration near the stacking fault with its local SFE, realizing the point-to-point predictions.

Inspired by the correlation of SFE to the local concentration, we plot the SFE against the number of Mo/Ti/V/Nb/Ta in the 1st-near SF (the 1st-near SF refers to two layers of atoms involved in the stacking fault) in Fig. 1 and S1. The increase of Mo (i.e., Mo aggregation) raises SFE, while the increase of Ti (i.e., Ti aggregation) lowers SFE just like the conventional Suzuki effect. For the correlation, Mo is the strongest, followed by Ti (see Fig. 1), and the other three (i.e., V, Nb, and Ta) are the weakest (see Fig. S1). These results demonstrate the predominant role of Mo and Ti concentrations in determining the SFE. We manually construct the 1st-near SF with high Mo content, and good linearity still exists for the Mo and Ti, further manifesting their predominant role (see Fig. S2). When extended to the 2nd-near SF, however, the linearity significantly decreases for Mo but slightly increases for Ti (see Fig. S3), which means that the effect of Mo is more localized than Ti. Nevertheless, the SFE is widely distributed at a certain Mo concentration, which is more prominent for Ti. The results further identify the literature findings that the correlation between SFE and an individual concentration fails to estimate SFEs quantitatively(*15, 27*).

Considering the opposite trend of Mo and Ti concentrations with SFE, we test four formulations in the TiVNbTaMo system by including the 1st- and 2nd-near Mo/Ti concentrations (see Fig. S4 and the details in Supplementary Note 2). We propose the descriptor—local concentration ratio ($D_{lcr}$), which is settled as follows:

$$D_{lcr} = \frac{\sum C_{VEmax,1}}{\sum C_{VEmin,1} + 0.5 \sum C_{VEmin,2}} \tag{1}$$

where the $C_{VEmost/VEleast,1/2}$ is the concentration of elements with the most/least valence electrons within the 1st-/2nd-near SFs (i.e., Mo and Ti for TiVNbTaMo HEAs). The 2nd-near SF refers to the two layers that extend outward from the 1st-near SF(*5*). The weighting factor 0.5 means that for the element with the least valence electron, the concentration of the 2nd-near SF ($C_{VEmin,2}$) has a smaller contribution compared with that of the 1st-near SF ($C_{VEmin,1}$), but is indispensable for quantitative prediction (see Fig. S4 and the details in Supplementary Note 2). As a result, the SFE shows a rather good linearity with the proposed descriptor $D_{lcr}$ (see Fig. 2(a) and the details in Supplementary Note 3). The MAE is ~39.02 mJ/m$^2$, with the $R^2$ as high as 0.820, in contrast to the 0.676 and 0.351 for Mo and Ti concentration, respectively (see Fig. 1(a) and (b)). Notably, this is the first time that a single descriptor, based on the local concentration, realizes the point-to-point quantitative predictions for SFEs.

To identify the role of different elements in the SFE of refractory HEAs, we adopted the three strategies: substituting one element with another; removing one element; changing the atomic proportion (i.e., deviating from equimolar).

The proposed descriptor $D_{lcr}$ is still applicable to determine SFE when Mo of TiVNbTaMo is substituted by W—i.e., TiVNbTaW, or when Ti of TiVNbTaMo is substituted by the Zr/Hf—i.e., ZrVNbTaMo, and HfVNbTaMo (see Fig. 2(a) and (b)). These results demonstrate the elements with the same number of valence electrons (namely in the same group) follow a similar rule in determining SFE. When substituting V with the W in the TiVNbTaMo, Mo and W with the identical number of valence electrons still play a similar role and work together in an additive way to determine SFE. The results of substitutions demonstrate that the number of valence electrons plays an essential role in predicting SFEs of HEAs: the elements with the most and least valence electrons in HEAs predominate.

We remove one element in the above quinary HEAs (TiVNbTaMo, TiVNbTaW, ZrVNbTaMo, HfVNbTaMo, and TiNbTaMoW): the first one is to remove one V B element (i.e., V, Nb, and Ta; labeled as Group I); the second is with IV B elements absent (i.e., Ti, Zr, and Hf; labeled as Group II); the third is with VI B elements absent (i.e., Mo, and W; labeled as Group III). For Group I, removing V from TiVNbTaMo to obtain TiNbTaMo, $D_{lcr}$ still linearly scales with SFE (see Fig. 2(c)), with $R^2$ up to 0.84. The similarity further demonstrates that the elements with the number of valence electrons between the most and least play a negligible role. However, removing Ti from the TiVNbTaMo, TiVNbTaW, and TiNbTaMoW to obtain VNbTaMo, VNbTaW, and NbTaMoW (see Fig. 2(c) and (d)), the V, Nb, and Ta show the least valence electrons and thus take the role of former Ti in determining SFE, entering the denominator of the $D_{lcr}$ expression. When removing Mo from ZrVNbTaMo and HfVNbTaMo to obtain ZrVNbTa and HfVNbTa (see Fig. 2(e)), the V, Nb, and Ta have the most valence electrons, taking the role of former Mo and entering the numerator of the $D_{lcr}$ expression. These results demonstrate that for a given element in HEAs (such as V), its influence on SFE strongly depends on the elemental composition of HEAs: when it has the most valence electrons, it increases SFE; when it has the least, it decreases SFE; however when it has the number of valence electrons between the most and least, it contributes slightly to SFE.

We also adopt the non-equimolar quaternary HEAs, such as TiNbTaMo2 and TiNbTa2Mo (see Fig. 2(f)). The descriptor $D_{lcr}$ linearly scales with the SFE, like its equimolar counterpart TiNbTaMo,

What is obvious is that the variation in the proportion of elements changes the local concentration, as well as the slope and intercept of the linear relationship. However, it does not impair the good linearity between $D_{lcr}$ and SFE. All the findings demonstrate that our descriptor quantitatively captures the local concentration-based nature of the SFEs in the refractory HEAs and reveals the role of any given element in determining the SFEs.

In the complex HEAs, our results indicate that the number of valence electrons still serves as an important parameter in distinguishing the difference between elements, just like in binary dilute alloys, but the intrinsic determinant of SFEs is the corresponding elemental concentration, rather than the number of valence electrons. In dilute binary alloys, such as Nb-(*35*) and Mg-based(*36–38*) alloys, the trend of SFE has been correlated with the valence electron difference or size difference compared with the matrix element. This is attributed to that the denser charge density with more valence electrons means stronger bonding and thus hinders the formation of SF by shearing(*36*). In contrast, the size difference has little effect on the SFE of HEAs: as shown in Fig. 2(b), the Zr and Hf atoms have larger atomic radii than others by about 8%~20% (according to the atomic radius listed in Ref. (*28*)), but the scaling relation between SFE and $D_{lcr}$ keeps good when substituting Ti with Zr or Hf (see Fig. 2(a) vs. Fig. 2(b)).

**The physical origin of the model.** We try to understand the physical origin of the descriptor $D_{lcr}$ with the cohesive energy $E_{coh}$ (that is essentially determined by the *d*-band width $W_d$): SFE ~ $D_{lcr}$ ~ $E_{coh}$. Intuitively, the strong cohesion means the large resistance of deformation to form a stacking fault, leading to SFE $\propto E_{coh}$. A good scaling relation of SFE and $E_{coh}$ has been found in hcp metals(*39*). We demonstrate that the scaling relation still exists for all TMs splitting into the two groups—one is the early transition metals (TMs), and the other is the rest removing the Cr, Fe, Mn, Tc, and Re outliners (see Fig. 3(a) and (b); Fig. S6). Moreover, in γ-Fe-Mn and Ti-Al alloys, the cohesive energy follows the same trend as SFE towards the change of concentration, further demonstrating the correlation between $E_{coh}$ and SFE(*40*, *41*).

Then, we try to understand why the elements with the least valence electrons play a more delocalized role than those with the most valence electrons. According to the TB model(*42*, *43*), the *d*-electrons mainly control the variation of the cohesive energy from one TM to the next, while the contribution of *s*-electrons to the cohesive energy ($E_{coh,s}$) is small and can be approximately considered constant due to the half-filled *s*-band filling. However, when including $E_{coh,s}$ via the free electron approximation in the Friedel model, the results show that $E_{coh,s}$ follows the same trend as the total $E_{coh}$ for the early TMs—i.e., Sc/Y/La, Ti/Zr/Hf, and V/Nb/Ta. This indicates that the *s*-electrons play an important role in determining the trend of $E_{coh}$ for the early TMs(*43*). Accordingly, for early TMs, $E_{coh,s}$ exhibits much better linearity with SFE than $E_{coh,d}$ (see Fig. 3(a) vs. S6(b) and (c)). There is no surprise that for the late TMs, the SFE scales well with $E_{coh,d}$ but not $E_{coh,s}$ (see Fig. 3(b) vs. S6(d) and (f)). These results demonstrate that the *s*-electrons play an important part in determining SFE for Ti, Zr, and Hf (see Fig. 3(a) and S6(a)-(c)), while the *d*-electrons predominate in Mo and W (see Fig. 3(b) and S6(e)-(f)). Since the *s*- and *d*-electrons show nonlocal and local characteristics respectively, the influence of Ti, Zr, and Hf on the bonding of HEAs is more delocalized than that of Mo and W. Therefore, the descriptor $D_{lcr}$ includes the concentration effect of both the 1st- and 2nd-SFs of Ti/Zr/Hf but only that of the 1st-SF of Mo/W.

Finally, we try to understand why the number of valence electrons serves as a criterion to select elements for including their concentrations in the descriptor $D_{lcr}$. This can be rationalized by the change in the density of states (DOS)—such as $W_d$ and $N_d$—from TMs to perfect HEAs to stacking-faulted HEAs. When alloying, the elements have to reshape to adapt to a common broadening: the elements with narrow broadening become less localized, while the elements with wide broadening become less delocalized. A typical example is that the DOS of *d*-bands for binary MoPd alloys is similar to that of Ru—the average element corresponding to MoPd(*44*). Besides, if forming binary alloys from TMs, the difference in $W_d$ predominates in the mixing enthalpy, while the electron transfer contributes much small(*42*, *43*, *45*, *46*). The conservation of *d*-band filling ($N_d$) has been demonstrated in the bimetallic surfaces, unaffected by the strain and ligand effects(*47*). Inspired by the physical picture in binary alloys, we thus assume that the $W_d$ of each component in HEAs approaches the average of its value in the pure phase (see Fig. 3(c) vs. (d)), but each component maintains the *d*-band filling ($N_d$) of its pure phase. The results of TiVNbTaMo fulfill this assumption that the *d*-band DOS of each component adapts to a nearly common broadening, but the *d*-band filling of each component slightly changes compared with that in the pure phase (see Fig. 3(c) and (d)). When forming the stacking fault from the prefect HEAs, the change in both the $W_d$ and $N_d$ is even smaller (see Fig. 3(d) and (e); Fig. S7). Overall, the *d*-band width $W_d$ becomes average from TMs to HEAs but slightly changes from perfect to stacking-faulted HEAs, while the *d*-band filling $N_d$ always maintains from TMs to perfect HEAs to stacking-faulted HEAs. These results may explain why the role of a given element in determining SFE strongly depends on its valence-electron number (i.e., $N_d$) relative to other components, instead of the *d*-band width (i.e., $W_d$).

Apart from the clear physical picture, our model also shows the ability to point-to-point quantitative predictions. The MAE of all systems averages at 27.73 mJ/m$^2$, with the error below ~4% in the range of SFE between 400~1100 mJ/m$^2$. The MAE lies between 5.21 and 49.71 mJ/m$^2$, with the error of each HEA system ranging from 6.23% to 14.18%. For the TiNbTaMo, MoNbTaW, Mo2NbTaTi, and MoNbTa2Ti systems, the MAEs are excellently low (i.e., 7.76, 7.24, 8.26, and 5.21 mJ/m$^2$). For the systems with large MAEs, such as TiVNbTaMo (39.02) and TiNbTaMo (33.73), the large MAE is due to the use of small supercells in DFT calculations, which can be greatly relieved by using larger supercells with more atoms(*5*, *12*, *16*, *48*). The MAE reduces significantly from 39.02 (60 atoms) to 22.40 (120 atoms) for the TiVNbTaMo HEAs, and from 33.73 (60 atoms) to 7.76 (960 atoms) for the TiNbTaMo HEAs (see Fig. 3(f) and the details in Supplementary Note 3).

Notably, the local concentration-based descriptor $D_{lcr}$ that realizes the point-to-point quantitative prediction further demonstrates the local property of SFE. In contrast, the SRO and CW based on the global properties show an intrinsic drawback in accounting for the SFE and thus are limited to qualitative correlation. Most importantly, our model is compatible with the conventional Suzuki mechanism (that is assumed to be out of data in HEAs) in determining the SFE of HEAs. Besides, the involved parameters of our model are easily accessible, and thus the model is more convenient for realistic applications.

**Discussion**

In summary, we proposed an effective descriptor for determining the SFE of refractory HEAs, using the local concentration ratio of components with the most different number of valence electrons. The descriptor correlates with the individual local SFE and realizes the point-to-point quantitative prediction for the SFEs of refractory HEAs. We find that for a given HEA, the elements with the most and least valence electrons play a determining role in the SFEs, while the elements in between have little contribution. The elements with the least valence electrons act in a more delocalized way than the elements with the most valence electrons because the former's *s*-component of cohesive energy plays a part in their bonding variation. Notably, our model rebuilds the framework of the conventional Suzuki mechanism (that is generally considered invalid in HEAs) in determining the SFE of HEAs, and its local nature essentially rules out the models based on the global SRO and CW. Our scheme uncovers the electronic origin, the coupling rule of different components, and the intrinsic determinants for the SFE of HEAs, all of which build a novel physical picture for understanding the SF and SFE of HEAs and provides an effective tool for predicting the SFE of HEAs.

## Methods

All the bcc refractory HEA structures are generated with the special quasi-random structure (SQS) method in the Alloy Theoretic Automated Toolkit (ATAT) code(*49*). The default supercells contain 60 atoms. We test the convergence of SFE to the size of the supercell by using 120-atom supercells (see the details in Supplementary Note 1). The super large supercells with 960 atoms are also used to simulate the realistic HEAs with the possible randomness of components(*48*). Besides, the calculations for all transition metals (TMs) are conducted with their most stable crystal structures—i.e., bcc, fcc, and hcp.

The density functional theory (DFT) calculations with the 60- or 120-atom supercells are performed in VASP code with the recommended PAW potentials(*50*, *51*). The same energy cutoff of 400 eV is used throughout for the comparison of different HEA systems. The geometric relaxation is performed at the PBE level of theory(*52*), converged within $10^{-6}$ eV and 0.02 meV/Å for the electronic and ionic relaxation, respectively. The first-order Methfessel Paxton smearing of 0.2 eV and the $R_k$ length for gamma-centered k-point meshing at ~30 Å is used to get well-converged results. The computations with the 960-atom supercells are conducted with the MTP machine learning force field (MLFF) implemented in the lammps code (*27*, *53*, *54*).

## Acknowledgments

**Funding:** The authors are thankful for the support from the National Natural Science Foundation of China (Nos. 22173034, 11974128, 52130101), the Opening Project of State Key Laboratory of High Performance Ceramics and Superfine Microstructure (SKL201910SIC), the Program of Innovative Research Team (in Science and Technology) in University of Jilin Province, the Program for JLU (Jilin University) Science and Technology Innovative Research Team (No. 2017TD-09), the Fundamental Research Funds for the Central Universities, and the computing resources of the High Performance Computing Center of Jilin University, China.

**Author contributions:** W.G. and Q.J. conceived the original idea and designed the strategy. C.M. performed the DFT calculations. W.G. derived the models and analyzed the results with the contribution from C.M.. C.M. and W.G. wrote the manuscript. C.M. prepared the Supplementary Information and drew all figures. All authors have discussed and approved the results and conclusions of this article.

**Competing interests:** The authors declare no competing interests.

## Data availability
All the data related to this work are available from the authors upon reasonable request

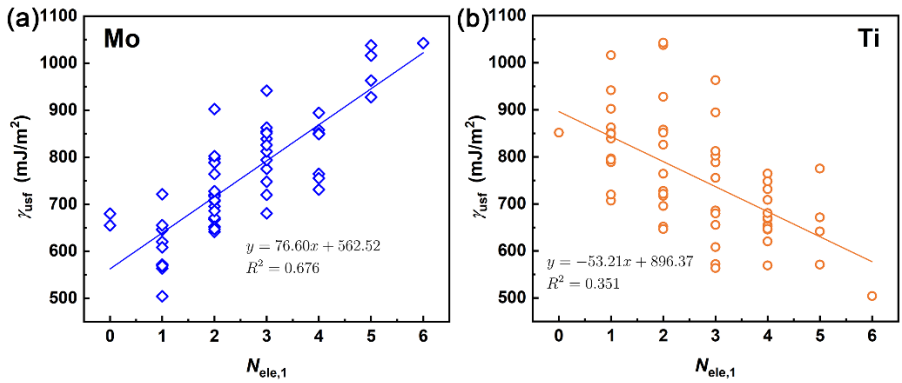

**Fig. 1 The correlation of stacking fault energy (SFE) with individual elemental concentration.** The $\gamma_{usf}$ is a function of the number of elements in the 1st-near stacking fault ($N_{ele,1}$): Mo (**a**) and Ti (**b**). The results for V, Nb, and Ta are shown in Fig. S1.

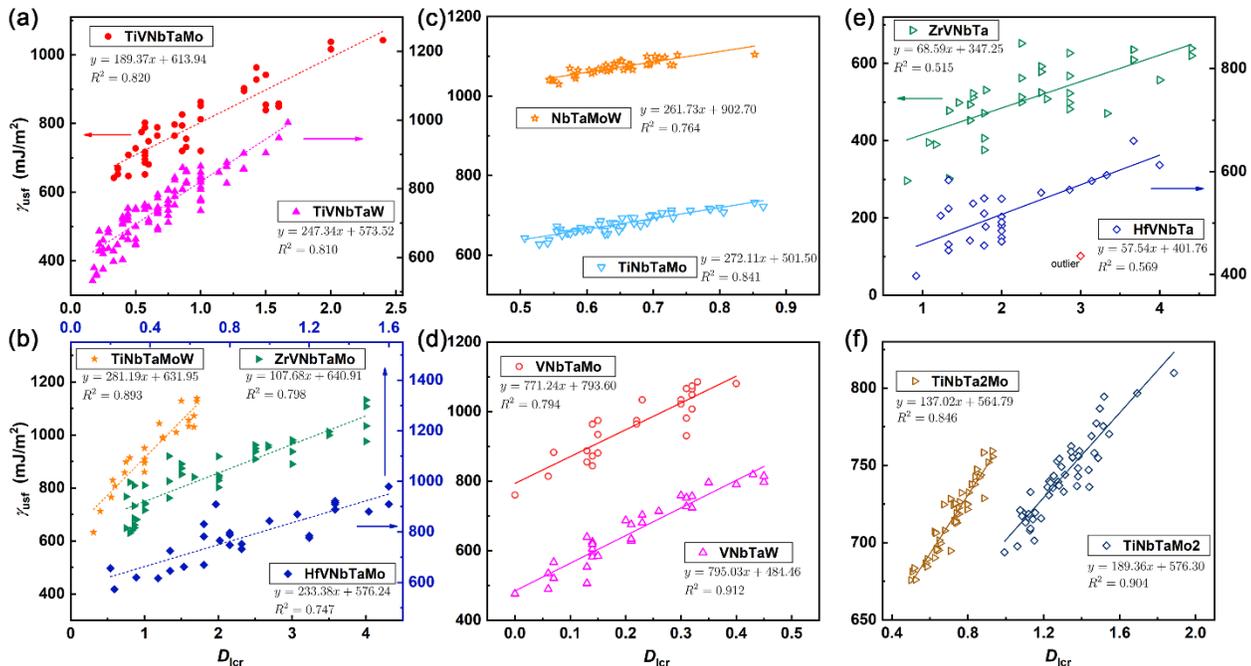

**Fig. 2 The stacking fault energy (SFE) against the descriptor $D_{lcr}$ in the refractory HEAs.** (**a**) TiVNbTaMo and TiVNbTaW. (**b**) ZrVNbTaMo, HfVNbTaMo, and TiNbTaMoW. (**c**) TiNbTaMo (Group I) and NbTaMoW (Group II). (**d**) VNbTaMo and VNbTaW (Group II). (**e**) ZrVNbTa and HfVNbTa (Group III). (**f**) Non-equimolar TiNbTaMo2 and TiNbTa2Mo. All the X (top-X and bottom-X) and Y (top-Y and bottom-Y) axes refer to the $D_{lcr}$ and $\gamma_{usf}$, respectively. The results in (**c**) and (**f**) are from the MTP MLFF with the 960-atom supercells(*27, 53*), while the rest are from the DFT method with the 60-atom supercell.

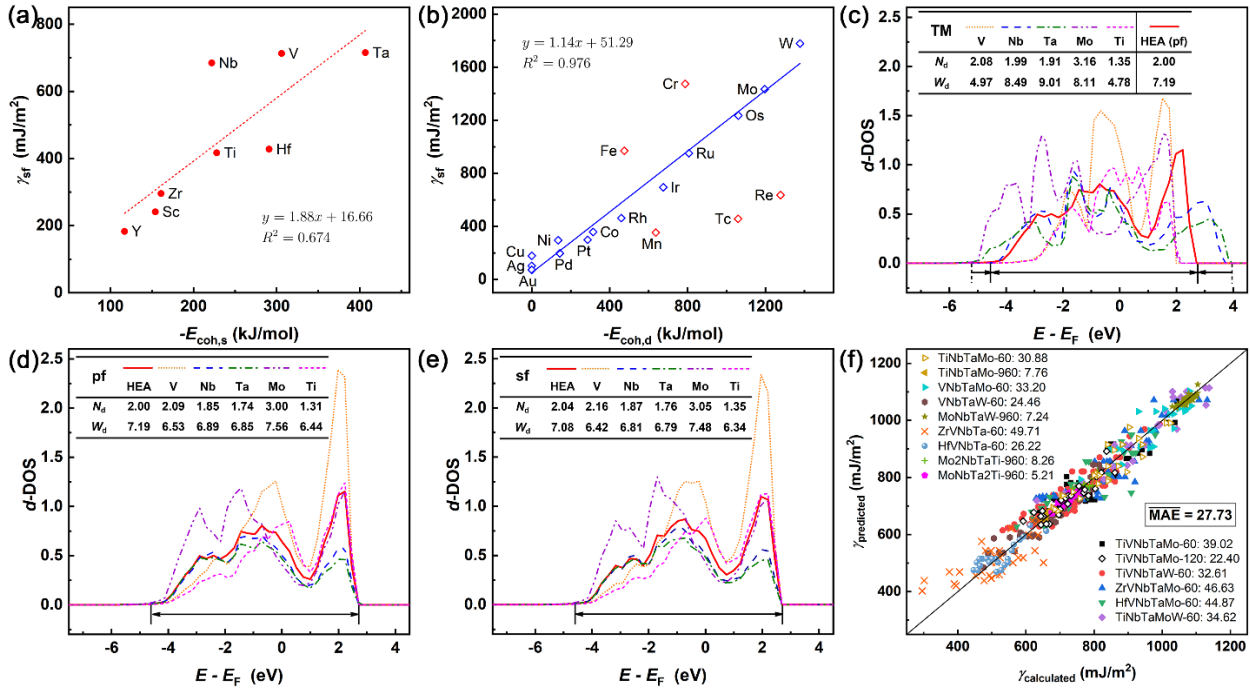

**Fig. 3 The physical origin of $D_{lcr}$ as well as its quantitative predictions.** (**a**) and (**b**) The stacking fault energy (SFE; $\gamma_{sf}$) against the contribution of *s*- and *d*-electrons to the cohesive energy ($E_{coh,s}$ and $E_{coh,d}$) for the early (**a**) and late (**b**) transition metals (TMs) (see also Fig. S6). The Cr, Fe, Mn, Tc, and Re are outliers. Data of $E_{coh,s}$ and $E_{coh,d}$ are from Ref. (*43*). (**c**) The comparison of the density of states (DOS) for *d*-bands (*d*-DOS) between TMs (i.e., V, Nb, Ta, Mo, and Ti) and perfect (pf) TiVNbTaMo HEA. (**d**) The comparison of the *d*-DOS between perfect (pf) TiVNbTaMo HEA and its components (i.e., V, Nb, Ta, Mo, and Ti). (**e**) The comparison of the *d*-DOSs between stacking-faulted (sf) TiVNbTaMo HEA and its components (i.e., V, Nb, Ta, Mo, and Ti). In (**c**), (**d**), and (**e**), the *d*-band filling ($N_d$) and width ($W_d$) per atom are listed. The numerical values of TMs are rational, following a similar trend as those obtained by the Friedel model (see Fig. S7(a) and (b)). To ensure high comparability, the same integration interval and processing method are used from the TMs to perfect (pf) to stacking-faulted (sf) HEAs. Apart from the similarity of $N_d$ and $W_d$, the similarity in the shape of *d*-DOS between perfect and stacking-faulted HEAs in (**d**) and (**e**) is further illustrated in Fig. S7(c). (**f**) Comparison between the predicted SFE ($\gamma_{predicted}$) and the DFT/MLFF calculated ($\gamma_{calculated}$) values for the quinary and quaternary HEAs. The 60, 120, and 960 refer to the number of atoms in the supercell. The MAEs for each system are presented in the legend. The average MAE for all systems is 27.73 mJ/m$^2$.

# Supplementary Materials for

## A Local Concentration-based Descriptor Predicting the Stacking Fault Energy of Refractory High Entropy Alloys

Cong Ma *et al.*

*Corresponding author. Email: wgao@jlu.edu.cn; jiangq@jlu.edu.cn.

**This PDF file includes:**

    Supplementary Text
    Figs. S1 to S7
    References (1 to 10)

# Supplementary Text

## Supplementary Note 1: computational details

All the intermediate structures in the evolution of special quasi-random structure (SQS(55)) generation are kept for density functional theory (DFT) calculations for the diversity of structures (i.e., the local environment in the vicinity of stacking fault). The most possible slip systems for bcc crystals are <111>{110}(56, 57). Therefore, the supercells with 60 atoms are reorientated along [111]×[11-2]×[-110] with respect to the conventional supercell(57). As a result, the {110} slip plane is parallel to the *x-y* plane, and the <111> slip direction is along the *x* direction. The supercell in the *z* direction is large enough to avoid spurious interactions between the stacking faults. We also adopted larger supercells with 120 atoms to test the convergence. To simulate the realistic HEAs, we used the super large supercells with 960 atoms (i.e., 4[111]×4[11-2]×5[-110]) to simulate the possible randomness of components(58).

The shear method is used to generate stacking-faulted structures, without introducing a vacuum (compared with the slab method (59)) and with fewer atoms in the supercell(57, 60). The optimal lattice parameters are fitted according to the equation of state (EOS). With the optimal lattice, all the atoms are only allowed to relax normally to the {110} plane with the supercell fixed. The results with the constrained relaxation must lie in between the simple shear (without relaxation) and pure shear (relaxing both atomic position and cell)(61).

The generalized stacking fault energy (GSFE) is derived from the energy cost per area when shifting one part of the crystal against the other part in the slip plane(62, 63):

$$\gamma_{sf} = \frac{E_{sf} - E_{pf}}{A}$$

where $E_{sf}$ and $E_{pf}$ are the total energies of crystals with and without stacking fault, and *A* is the stacking fault area.

## Supplementary Note 2: test for the descriptor's formulation

Due to the opposite correlation of Mo and Ti concentrations with SFE, we test four different formulations for the TiVNbTaMo system (see Fig. S4). We can see that further including the Mo atoms of the 2nd-near SF always leads to poor linearity (see Fig. S4(c) vs. (d)). However, further including the Ti of the 2nd-near SF improves the linearity (see Fig. S4(a) vs. (b)).

We also test the weighting factor of Ti content in the 2nd-near stacking fault. The fitting coefficient $R^2$ peaks at the weighting factor of 0.5, demonstrating that the 2nd-near concentration effect of Ti has a small contribution than the 1st-near effect, but plays an important part in the quantitative prediction for both TiVNbTaMo and TiVNbTaW (see Fig. S4(e) vs. (f)).

## Supplementary Note 3: supercell size convergence

We also adopted larger supercells with 120 atoms to test the convergence of SFE with respect to $D_{lcr}$. For the template TiVNbTaMo HEAs, the similar linearity between the 60-atom and 120-atom supercells demonstrates that the results have converged with respect to the sizes of the supercell (see Fig. S5(a)). For the TiNbTaMo HEAs, using a small 60-atom supercell instead of a 960-atom increase the MAE significantly, but the fitting coefficients ($R^2$) are maintained (see Fig. S5(b)).

**Fig. S1.**

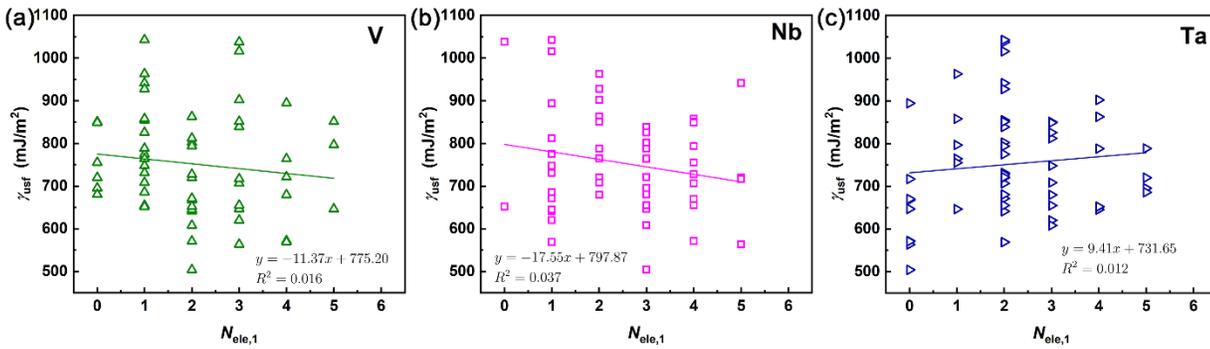

**Fig. S1 The correlation of stacking fault energy (SFE) with individual elemental concentration.**
The $\gamma_{usf}$ is a function of the number of elements in the 1st-near stacking fault ($N_{ele,1}$): V (**a**), Nb (**b**), and Ta (**c**).

**Fig. S2.**

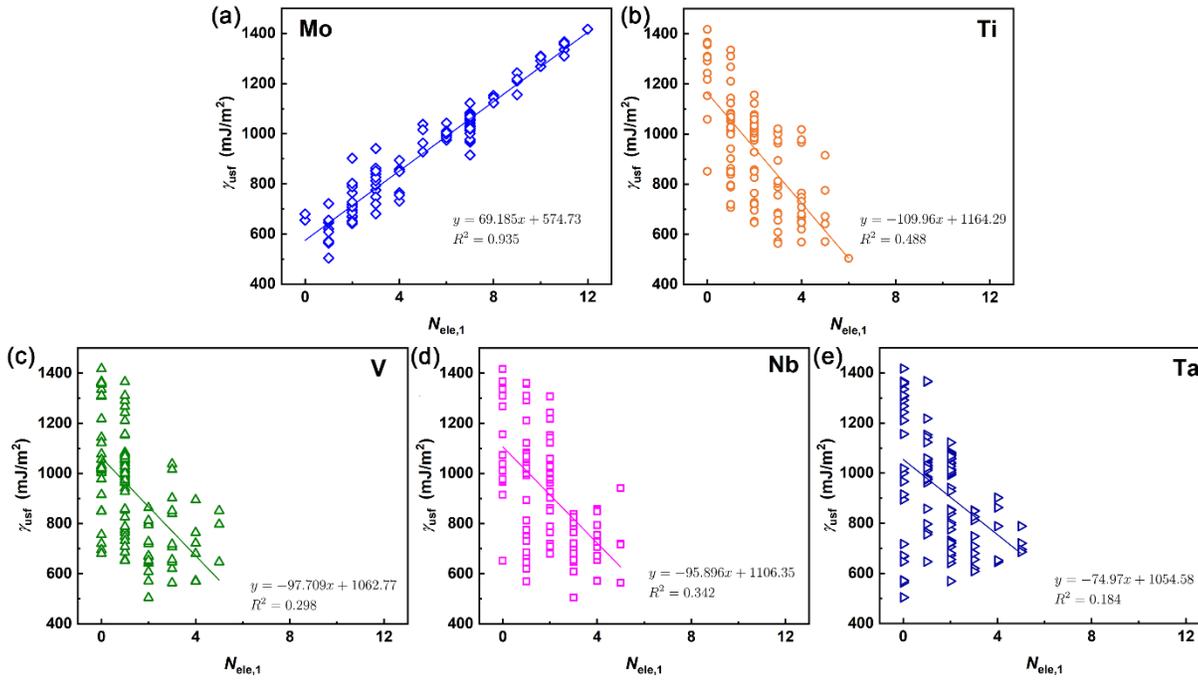

**Fig. S2 The correlation of stacking fault energy (SFE) with individual elemental concentration.** For the manually-constructed 1st-near stacking fault with high Mo content, the $\gamma_{usf}$ as a function of the number of elements in the 1st-near stacking fault ($N_{ele,1}$): Mo (**a**), Ti (**b**), V (**c**), Nb (**d**), and Ta (**e**).

**Fig. S3.**

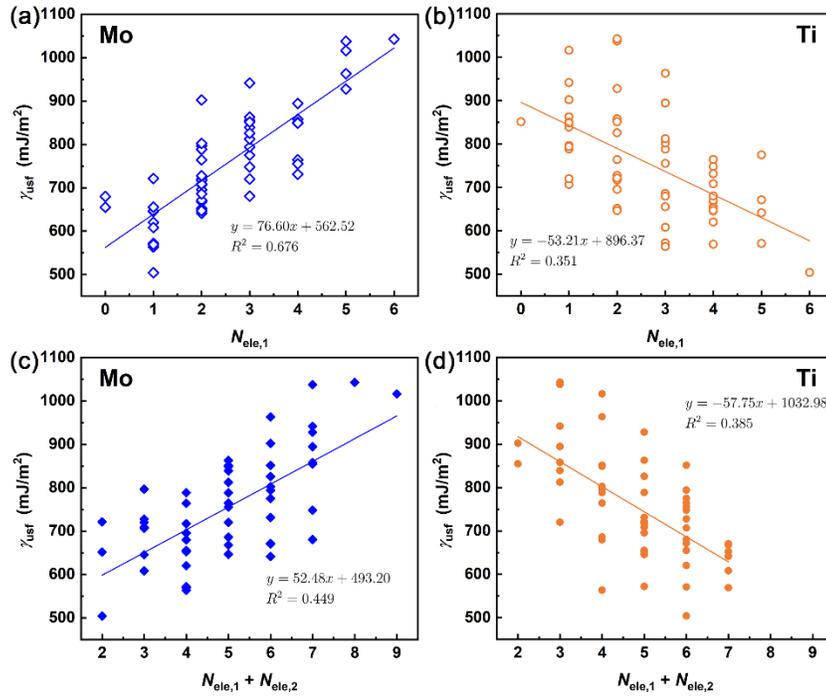

**Fig. S3 The correlation of stacking fault energy (SFE) with individual elemental concentration.**
(**a**) and (**b**) The $\gamma_{usf}$ as a function of the number of elements in the 1st-near stacking fault ($N_{ele,1}$): Mo (**a**) and Ti (**b**). (**c**) and (**d**) The SFE as a function of the number of elements in both the 1st- and 2nd-near stacking faults ($N_{ele,1} + N_{ele,2}$): Mo (**c**) and Ti (**d**).

**Fig. S4.**

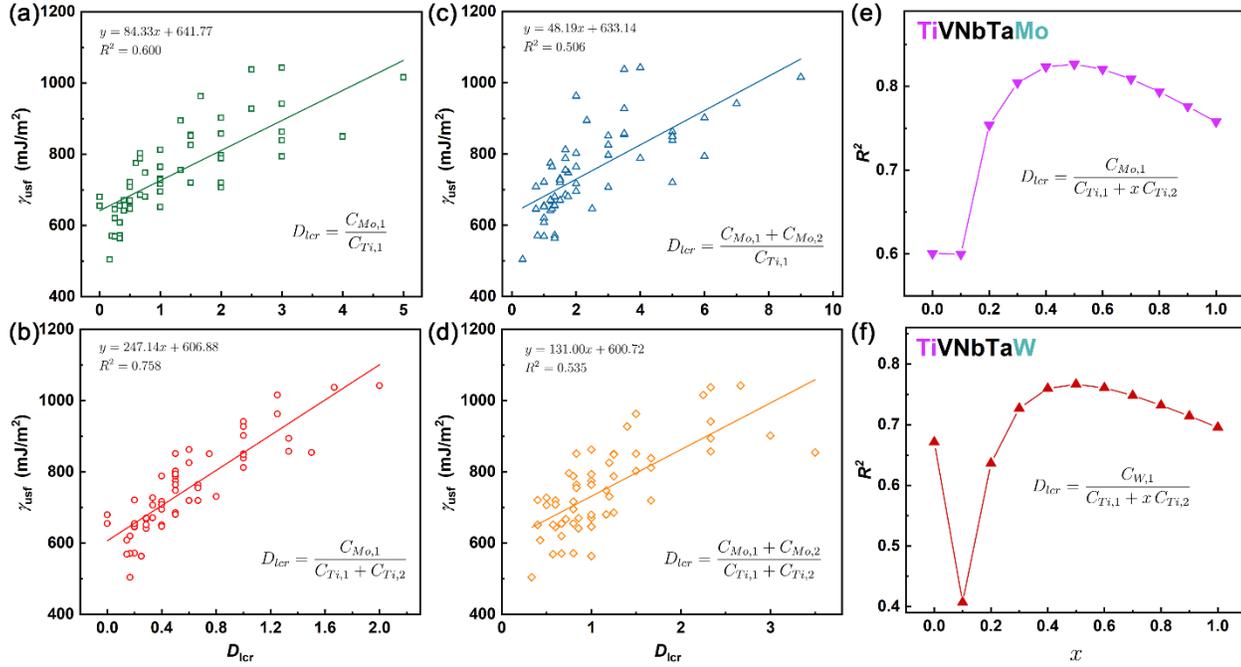

**Fig. S4 The stacking fault energy (SFE) against the descriptor $D_{lcr}$ with different formulations.** For TiVNbTaMo (**a**)-(**e**) and TiVNbTaW (**f**) HEAs, Mo and W show the most valence electrons, while Ti shows the least. The $C_{Mo/W/Ti, 1}$ and $C_{Mo/W/Ti, 2}$ refer to the concentrations of Mo/W/Ti atoms in the 1st- and 2nd-near stacking faults, respectively. (**e**) and (**f**) The fitting coefficient $R^2$ as a function of the weighting factor ($x$) for the $C_{Ti, 2}$. The $R^2$ peaks at the weighting factor of 0.5 for both the TiVNbTaMo (**e**) and TiVNbTaW (**f**), which means the 2nd-near concentration effect of Ti is small weighted but plays an important part in quantitatively predicting SFEs.

**Fig. S5.**

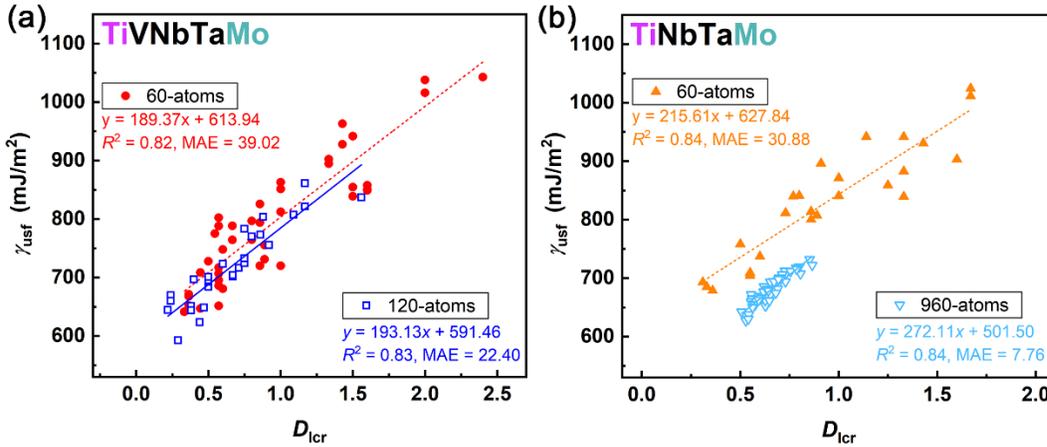

**Fig. S5 The stacking fault energy (SFE) against the descriptor $D_{lcr}$ in different sizes of the supercells.** (**a**) For the TiVNbTaMo HEAs with 60-atom and 120-atom supercells, the correlation between $\gamma_{usf}$ and $D_{lcr}$ is similar in the fitting coefficients ($R^2$), demonstrating that the good linearity has converged with respect to the sizes of supercell. The MAEs reduce significantly from small to large supercells. (**b**) For the TiNbTaMo HEAs with 60-atom and 960-atom in the supercell. The situation is similar to the TiVNbTaMo HEAs. The MAE greatly reduces using a 960-atom supercell.

**Fig. S6.**

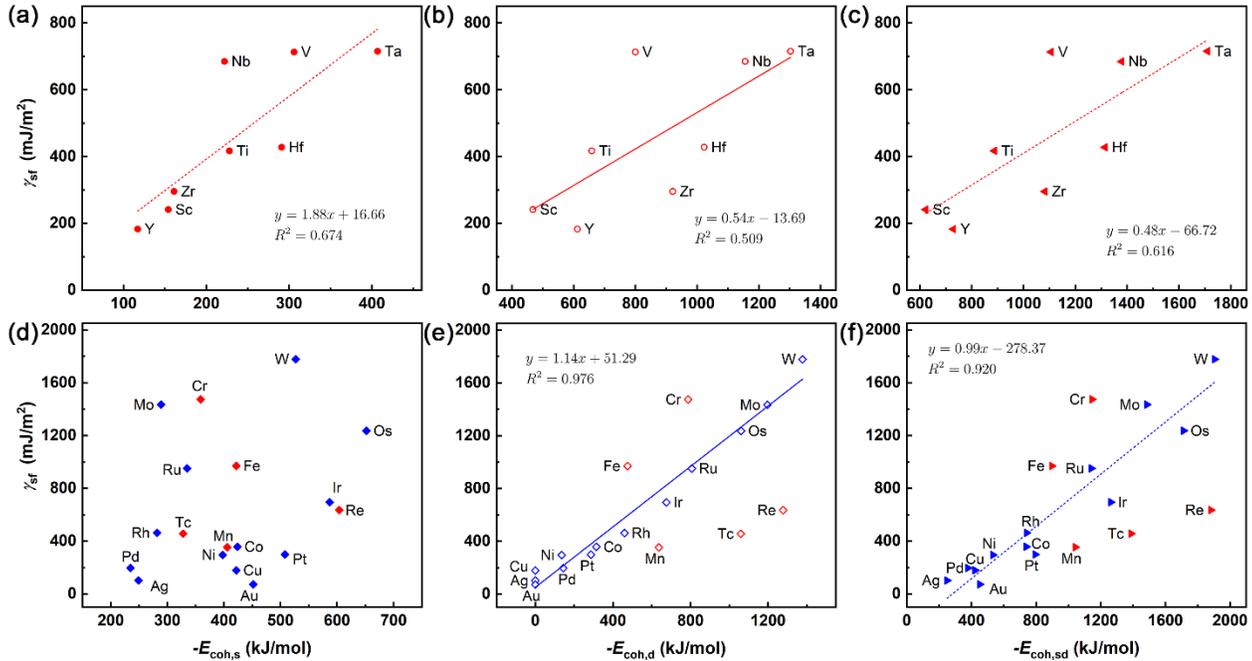

**Fig. S6 The stacking fault energy (SFE) against the cohesive energy ($E_{coh}$) for early and late transition metals (TMs).** The $\gamma_{sf}$ as a function of $E_{coh}$ for the early (**a**)-(**c**) and late (**e**)-(**f**) TMs. (**a**) and (**d**) the contribution of *s*-electrons to the cohesive energy ($E_{coh,s}$). (**b**) and (**e**) the contribution of *d*-electrons to the cohesive energy ($E_{coh,d}$). (**c**) and (**f**) the combination of the contribution of the *s*- and *d*-electrons to the cohesive energy ($E_{coh,sd}$). The cohesive energies are extracted from the literature(*64*). The $\gamma_{sf}$ values of TMs computed in this work are all based on the most stable crystal structures (i.e., bcc, fcc, and hcp). For the early TMs, the linearity between $\gamma_{sf}$ and $E_{coh}$ is much better for the $E_{coh,s}$ than $E_{coh,d}$ (see (**a**)-(**c**)). For late TMs, the SFE scales well with $E_{coh,d}$ except for the Cr, Fe, Mn, Tc, and Re outliers, while no rule exists for the $E_{coh,s}$ (see (**d**)-(**f**)).

**Fig. S7.**

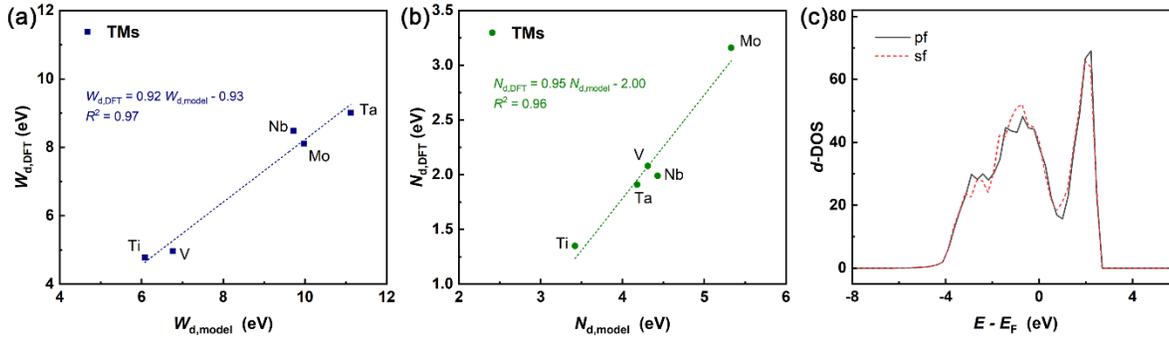

**Fig. S7 Validation of the results of the density of states (DOS) for the *d*-bands (*d*-DOS).** (**a**)-(**b**) Despite the discrepancy in numerical values, the *d*-band width ($W_d$) and filling ($N_d$) for transition metals (TMs) computed with the DFT method follow a similar trend as those based on the Friedel model(*64*). (**c**) The similarity in the shape of the *d*-DOS between the perfect (pf) and stacking-faulted (sf) TiVNbTaMo HEA (the total results for 60-atoms supercell).